\documentclass[a4paper]{article} 

\pdfoutput=1

\usepackage{hyperref}
\usepackage{url}
\hypersetup{
  pdfinfo={
    Title={Your Title Here},
    Author={Your Name Here},
    Subject={If you want to put something here, do so},
    Keywords={Add some keywords if you feel so inclined}
  }
}
\usepackage{pdfpages}
\usepackage{iclr2021_conference,times}
\usepackage{tabularx}

\usepackage{amsmath,amsfonts,bm}

\def\eqref#1{equation~\ref{#1}}

\def\1{\bm{1}}

\DeclareMathAlphabet{\mathsfit}{\encodingdefault}{\sfdefault}{m}{sl}
\SetMathAlphabet{\mathsfit}{bold}{\encodingdefault}{\sfdefault}{bx}{n}

\usepackage{graphicx}
\usepackage{xcolor}
\usepackage{xcolor,colortbl}

\newcommand{\mc}[2]{\multicolumn{#1}{c}{#2}}
\definecolor{Gray}{gray}{0.85}
\definecolor{LightCyan}{rgb}{0.88,1,1}
\usepackage{soul}
\usepackage{hyperref}
\usepackage{url}

\title{COVID-Net CT-S: 3D Convolutional Neural Network Architectures for COVID-19 Severity Assessment using Chest CT Images}

\author{Hossein Aboutalebi$^{1,3}$, Saad Abbasi$^{2}$, Mohammad Javad Shafiee$^{2,3}$, Alexander Wong$^{2,3}$ 

\\
$^1$Department of Computer Science, University of Waterloo, Canada\\
$^2$Department of Systems Design Engineering, University of Waterloo, Canada\\
$^3$Waterloo Artificial Intelligence Institute, University of Waterloo, Canada \\
\texttt{\{haboutal, srabbasi, mjshafiee, a28wong\}@uwaterloo.ca} 
}

\iclrfinalcopy 
\begin{document}

\maketitle

\begin{abstract}
The health and socioeconomic difficulties caused by the COVID-19 pandemic continues to cause enormous tensions around the world.
In particular, this extraordinary surge in the number of cases has put considerable strain on health care systems around the world.  A critical step in the treatment and management of COVID-19 positive patients is severity assessment, which is challenging even for expert radiologists given the subtleties at different stages of lung disease severity.  Motivated by this challenge, we introduce COVID-Net CT-S, a suite of deep convolutional neural networks for predicting lung disease severity due to COVID-19 infection.  More specifically, a 3D residual architecture design is leveraged to learn volumetric visual indicators characterizing the degree of COVID-19 lung disease severity.   Experimental results using the patient cohort collected by the China National Center for Bioinformation (CNCB) showed that the proposed COVID-Net CT-S networks, by leveraging volumetric features, can achieve significantly improved severity assessment performance when compared to traditional severity assessment networks that learn and leverage 2D visual features to characterize COVID-19 severity. 

\end{abstract}
\vspace{-0.1in}
\section{Introduction}
\vspace{-0.1in}
The coronavirus disease 2019 (COVID-19) pandemic has challenged the world at every level, from the health crisis to the social disruption and the economic collapse. Due to globalization and urbanization, the speed of the spread of the severe acute respiratory syndrome
coronavirus (SARS-CoV-2) that caused the COVID-19 pandemic has been continuously exceeding estimations around the world. 

Since March 2020, when the virus was for the first time declared a pandemic due to the death toll and the number of positive cases by the World Health Organization~\cite{jebril2020world}, the economic impacts of this disease has been enormous. Closure of small businesses, reduction in the workforce on the sites, trade disruption, and a sharp decline in the tourism industry are just a few early impacts of the economic consequences of this pandemic~\cite{pak2020economic}.
In this regard, the International Monetary Fund announced that the global economy was contracted by more than 3\% in 2020 due to the uncertainty, lower activity, and lack of productivity~\cite{brodeur2020literature}.

More importantly, from a global healthcare perspective, the impact of this pandemic has been devastating for healthcare systems around the world. The dramatic increase of patients with COVID-19 symptoms has overwhelmed hospitals and clinics.  As the demand has exceeded hospital capacities, problems ranging from the lack of PPEs and limited personnel to supply issues and increasing demand for ventilator machines and oxygen therapies have emerged.  As a result, healthcare providers are in a difficult situation to be selective in the admission process of patients~\cite{azoulay2020admission}.
A critical step in the COVID-19 clinical workflow is severity assessment, which provides clinicians with the information needed to support decisions related to the treatment and management of COVID-19 positive patients, ranging from Intensive Care Unit (ICU) admission to ventilator usage~\cite{tyrrell2021managing}. Providing a robust and consistent COVID-19 severity assessment system is therefore essential for boosting the quality of healthcare services and can potentially reduce the death toll caused by the pandemic.  However, assessing disease severity of COVID-19 positive patients is quite challenging even for expert clinicians given the subtleties at different stages of lung disease severity.  Motivated by this, we introduce COVID-Net CT-S, a suite of deep convolutional neural networks for predicting the lung disease severity due to COVID-19 infection based on computed tomography (CT) volumes.
\subsection{Related work}

From the early days of the pandemic, deep learning provided a new angle to overcome the pandemic challenges. Using deep convolutional neural networks (CNNs) as a screening tool for early detection of the disease has been a great area of research interesting. COVID-Net, which was proposed by \cite{wang2020covid}, was among the first screening tools which leveraged deep CNNs. In that work, CNNs were leveraged to analyze chest x-ray (CXR) images to distinguish COVID-19 infections from non-COVID-19 pneumonia and normal cases. \cite{jain2020deep} performs the same task on three different CNNs, including Inception network~\cite{szegedy2015going}, on the Kaggle repository of CXR images. In another work by \cite{sahlol2020covid}, a hybrid model was proposed combining a CNN with a swarm-based feature selection algorithm to increase detection accuracy. 

COVID-Net CT, which was proposed by proposed by \cite{gunraj2021covid}, leveraged computed tomography (CT) images to perform the same task. One advantage of COVID-Net CT is that it was trained on over a patient cohort of more than 4000 patients from different countries. In another work by  \cite{he2020sample}, the authors propose a sample efficient framework for diagnosing COVID-19.  They significantly leverage self-supervising transfer learning to avoid overfitting while getting high accuracy when the data is scarce.  In the proposed approach by   \cite{zhang2020clinically}, introduced a new dataset on CT images from Chinese hospitals as well as developed an AI system that comprised of an image segmentation algorithm followed by a classification neural network to detect COVID-19 cases. One of the main initiatives of this work is to utilize image segmentation to reduce the noise of the inputs for the classification task. 

There is also a stream of work focused on performing image segmentation to detect areas of COVID-19 infection. \cite{yan2004covid}  propose a deep learning framework for detecting affected ares in CT images. In particular, they propose a new feature variation block to elevate the segmentation performance by adaptively adjusting the features' global properties. \cite{muller2020automated},  propose a new segmentation framework to overcome small dataset problem. In particular, they leveraged a 3D U-Net combined with the generation of unique and random image patches for training datasets to further increase image segmentation's performance on CT images.

While there are numerous works on applying deep learning to screening and image segmentation, an area that is much less explored is deep learning for COVID-19 severity assessment. The seminal works in this area are COVID-Net S, a tailored deep convolutional neural network proposed by   \cite{wong2020covidnets} to predict geographic and opacity extent severity scores from CXR images, and the study by  \cite{cohen2020predicting} where they used a pretrained DenseNet \cite{huang2017densely} model to predict severity scores. In this regard, while it is common to use 2D architecture to handle CT images for COVID-19 detection like the works proposed in \cite{gunraj2021covid, zhang2020clinically}, in this work, we are motivated to leverage 3D architectures for the severity analysis to better capture the information across the depth dimension.

Here, the proposed COVID-Net CT-S networks leverage a 3D residual architecture design to predict COVID-19 severity based on CT volumes to better capture volumetric visual indicators characterizing the degree of COVID-19 lung disease severity. The proposed neural networks are trained and tested based on the patient cohort collected by the China National Center for Bioinformation (CNCB)~\cite{zhang2020clinically}, comprising of CT volumes and associated clinical severity scores. To illustrate the effectiveness of the proposed COVID-Net CT-S networks, we compare their performance to traditional severity assessment networks that leverage 2D visual features to characterize COVID-19 severity. The results show that taking advantage of volumetric information for characterizing COVID-19 infection characteristics can boost performance significantly. Furthermore, given the challenging nature of training 3D convolutional neural networks on volumetric CT scans captured under different settings and the imbalance nature of the patient cohorts, we describe a training regimen to improve generalization under such scenarios. 

\section{Methodology}

\begin{figure*}[!]
 \vspace{-0.5in}
\setlength{\tabcolsep}{0.01cm} 
\begin{tabular}{cc}
        \includegraphics[width=1\textwidth,height=0.35\textwidth]{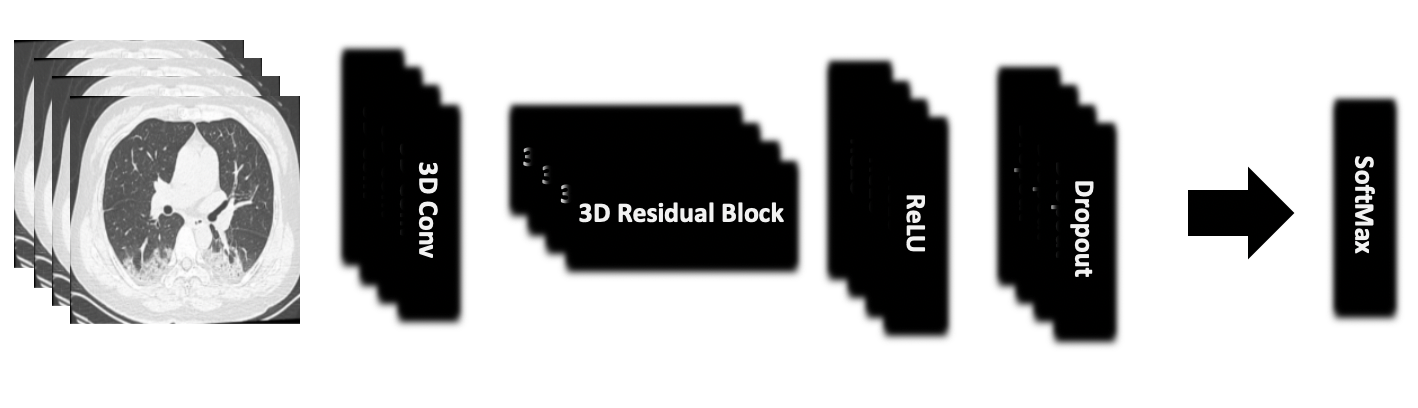}&
\end{tabular}
\vspace{-1cm}
\caption{Core architecture of the proposed COVID-Net CT-S for lung disease severity assessment of COVID-19 positive patients.}
\vspace{-0.2in}
\label{arc}
\end{figure*}

\textbf{Network Architecture.} The proposed COVID-Net CT-S architectures possess 3D residual deep convolutional neural network architectures to better capture subtle volumetric visual indicators characterizing COVID-19 lung disease severity.  More specifically, the three COVID-Net CT-S networks are comprised of 16, 33, and 50 3D convolutional layers, followed by the same number of 3D residual blocks for COVID-Net CT-S50, CT-S100, and CT-S152, respectively.  The output of the stack of 3D residual blocks is then fed into a ReLU layer, a dropout layer, and finally a softmax layer for making the final COVID-19 disease severity score (corresponding to low, medium, and high severity).     Figure~\ref{arc} illustrates the core architecture of the proposed COVID-Net CT-S neural network architecture. 
 
\begin{figure*}[h!]
\vspace{-0.3cm}
\hspace*{1.5cm}
\setlength{\tabcolsep}{0.01cm} 
\begin{tabular}{cc}
        \includegraphics[width=0.35\textwidth,height=0.35\textwidth]{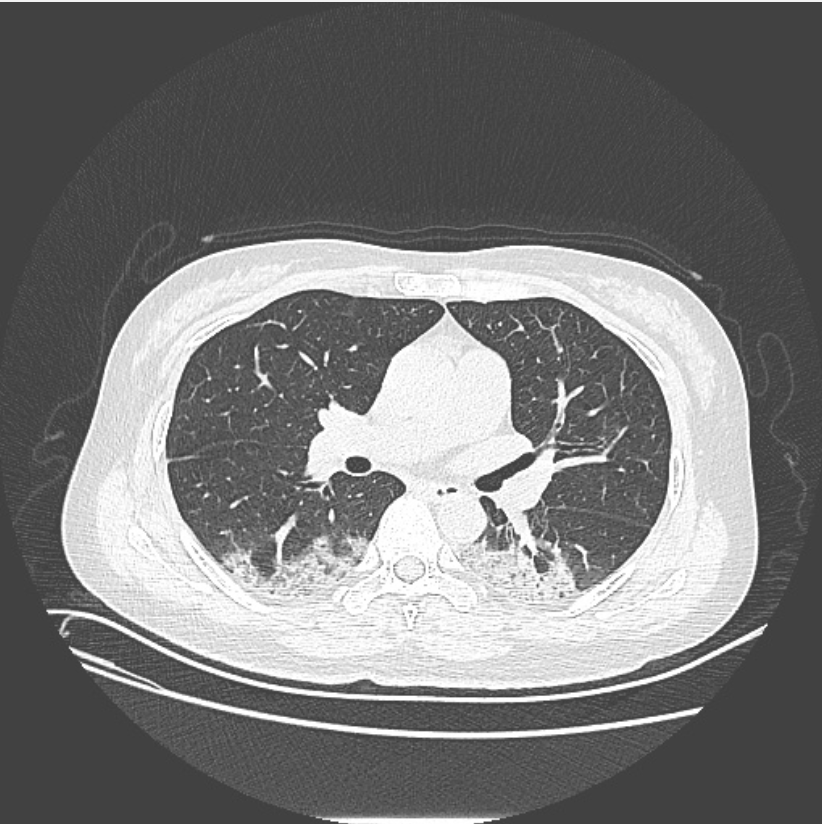}
		\includegraphics[width=0.35\textwidth,height=0.35\textwidth]{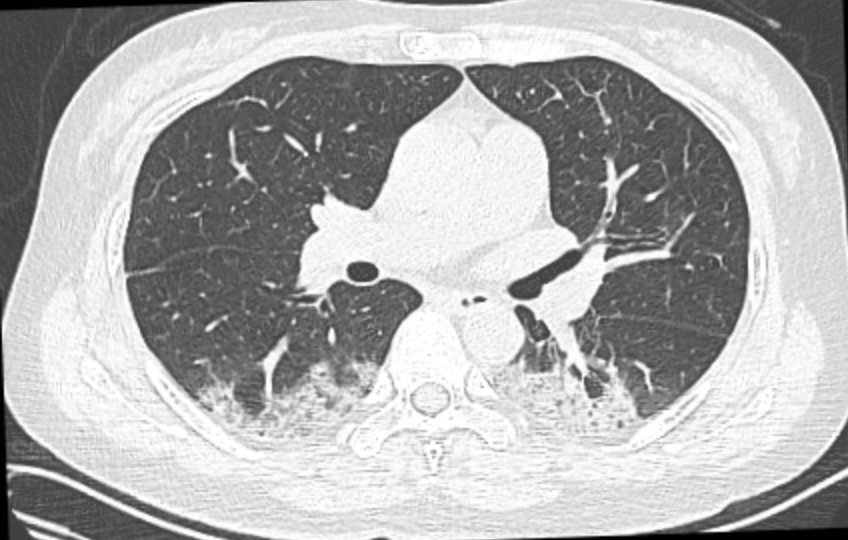}
		\\
		\hspace{-5cm} (a) Raw CT image  & \hspace{-4.5cm}(b) Post-Processed CT image
\end{tabular}
\caption{Comparison of CT image before (a) and after (b) preprocessing.}
\label{fig:acc-width}
\end{figure*}

\textbf{Training regimen.} Given the challenges of training such a 3D deep convolutional neural network on volumetric scans with high capture setup diversity, limited patient count, and highly imbalanced patient cohorts, we leveraged the following training regimen to greatly improve generalization: 
 \begin{itemize}
 \item {\bf Volumetric uniformization:}
 As mentioned earlier, CT images are captured using different CT scanners with different protocols, different hardware configurations, and the anatomical characteristics of patients (e.g., patient size) can vary greatly.  To alleviate this issue, we took inspiration from \cite{zunair2020uniformizing} and introduced a volumetric uniformization process based on spline interpolation to reconstruct all patient volumes in the cohort used in this study to a standardized volume depth of 40.
 \item {\bf Dropout Regularization:}
 Throughout extensive experiments, we observed that using dropout regularization can significantly improve severity assessment accuracy by as much as 10\%.  Therefore, we leverage dropout layers in all tested networks.
 \item {\bf Uniform Batch Sampling:}
As the severity levels amongst the patient cohort is highly unbalanced, we leverage a uniform batch sampling technique where a uniform number of samples from each severity level is drawn to construct a batch for each training iteration. This greatly increased the stability and convergence speed of the training process, as well as improved the generalization of the proposed networks.
 \item {\bf Preprocessing:}
 For preprocessing the CT volumes, we employed similar steps as those introduced in~\cite{gunraj2021covid}, but modified the cropping process to preserve more of the patient's body area while still maintaining patient table mitigation. Figure~\ref{fig:acc-width} shows the effect of the preprocessing step on an image slice from a CT volume.
 \end{itemize}

\vspace{-0.5cm}
\section{Experimental Results and Discussion}

\begin{figure*}[!]
\hspace*{3.5cm}
\setlength{\tabcolsep}{0.01cm} 
\begin{tabular}{cc}
        \includegraphics[width=0.4\textwidth,height=0.3\textwidth]{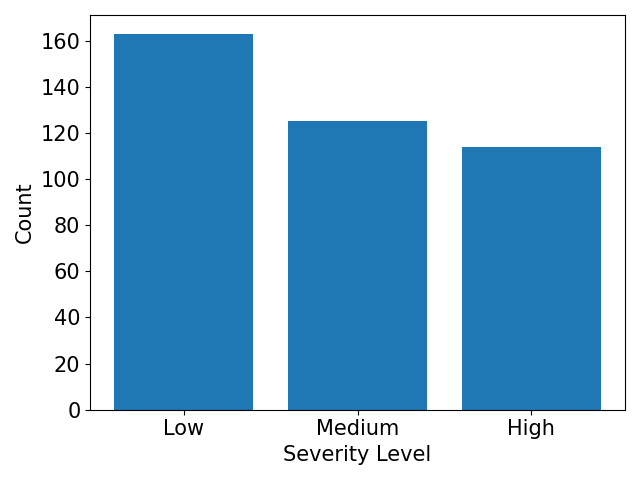}&
\end{tabular}
\vspace{0.2in}
\caption{Patient distribution for different severity levels in the CNCB dataset.}
\vspace{-0.05in}
\label{fig:histogram}
\end{figure*}

To evaluate the efficacy of the proposed COVID-Net CT-S networks, We performed all our experiments on the CNCB dataset \cite{zhang2020clinically}. In the dataset, there are a total of 267 patients with annotated severity analysis, and the resulting distribution of severity levels is shown in  Figure~\ref{fig:histogram}.  A total of 53 of the patient cases in the dataset are leveraged for testing and evaluation.  While the severity annotations have a dynamic range of 1 to 6, with 1 being least severe and 6 being the most severe, for the purpose of this study we have grouped them into 3 severity levels: 1) low (severity scores of 1-2), 2) medium (score of 3), and 3) high (scores of 4-6).  This was chosen given clinical similarities between these severity score groups in terms of the treatment and management regimen within these groups, and as such allows for clearer guidelines that link to the course of action to take.

One of the main architecture design choices for the proposed COVID-Net CT-S networks revolve around the hypothesis that leveraging of 3D convolutions can better characterize and learn the visual indicators associated with the degree of COVID-19 disease severity when compared to traditional severity assessment networks that leverage 2D convolutions.  Therefore, in this study we also trained a 50 layer residual network architecture comprised of 2D convolutions for performing COVID-19 severity assessment on a patient's condition by making predictions on all slices within a patient volume and performing majority vote on the predictions to determine the final severity level.

The performance of the tested models are evaluated based on the accuracy metric and is shown in Table~\ref{tab:res}. Furthermore,  Table 2 shows the confusion matrix for COVID-Net CT-S.  It can be observed from Table 1 that the proposed COVID-Net CT-S architectures outperform the traditional 2D architecture used in severity assessment networks by as much as 19.5\%, with the variant of COVID-Net CT-S sharing the same depth as the 2D architecture achieving 13\% higher accuracy. Furthermore, we can observe that using COVID-Net CT-S152 with dropout gives the best performance.  Based on the experimental results, it can be clearly observed that the proposed COVID-Net CT-S networks can achieve strong severity assessment performance when compared to traditional 2D severity assessment approaches.

\begin{table}[ht]
\parbox{.45\linewidth}{
\caption{Accuracy results on CNCB dataset}
\label{tab:res}
\begin{tabular}[t]{lcc}
\hline
Model&Accuracy\\
\hline
2D CT-S50~\cite{he2016deep} & 59.0 \\
COVID-Net CT-S152 & \textbf{78.5}\\
COVID-Net CT-S100 & 75.0\\
COVID-Net CT-S50 & 72.0\\
\hline
\end{tabular}
}
\hfill
\parbox{.5\linewidth}{
\label{tab:conf}
\caption{Confusion Matrix for COVID-Net CT-S }
\vspace{0.78cm}
\begin{tabular}{| c | c | c | c}
\hline
\rowcolor{LightCyan}
\mc{1}{} & \mc{1}{~~~Low~~~~} & \mc{1}{Medium} & ~~~High~~~ \\\mc{1}{~~~~Low~~~~} & 85\% & 8\% & 7\%
\\
\hline
\mc{1}{Medium} &10\% & 79\% & 11\% \\ 
\hline
\mc{1}{~~~High~~~~} &13\% & 15\% & 72\% \\ 
\end{tabular}
}
\end{table}
\vspace{-0.1in}
\section{Conclusion}
\vspace{-0.1in}
In this paper, we introduced COVID-Net CT-S, a suite of deep convolutional neural networks to predict the lung disease severity of COVID-19 positive patients based on CT image volumes.  While a 2D convolutional architecture has been one of the dominant architectures for COVID-19 assessment, in this paper we showed that for the case of severity assessment based on CT volumes, a 3D convolutional architecture could achieve improved prediction performance by better characterizing volumetric visual indicators associated with severity levels of COVID-19 infection.  Furthermore, we demonstrated that the use of a customized training regimen can improve the performance and generalization of networks when trained with a patient cohort with imbalanced severity levels and high equipment and protocol variance.

\bibliography{iclr2021_conference}
\bibliographystyle{authordate1}

\end{document}